\documentclass [a4paper, 11pt] {article}
\usepackage{amsmath}
\usepackage{amssymb}
\usepackage{amsfonts}
\usepackage{latexsym}
\usepackage{amsmath}
\usepackage{mathtools}
\usepackage{empheq}
\usepackage{graphicx}
\usepackage{caption}
\usepackage{subcaption}
\usepackage[toc,page]{appendix}
\usepackage{multicol}
\usepackage{color}
\begin{document}

\title{\textbf{Resonant Absorption of Kink MHD Waves in Inclined and Asymmetric Coronal Loops}}

\author{Sirwan Amiri$^1$\thanks{E-mail: asirwan9@gmail.com},
Kayoomars Karami $^1$\thanks{E-mail: kkarami@uok.ac.ir} and Zanyar Ebrahimi $^2$\thanks{E-mail: zebrahimi@maragheh.ac.ir}\\
$^1$\tiny{Department of Physics, University of Kurdistan, Pasdaran Street, P.O. Box 66177-15175, Sanandaj, Iran}\\
$^2$\tiny{Research Institute for Astronomy $\&$ Astrophysics of Maragha, University of Maragheh, P. O. Box 55136-553, Maragheh, Iran}}
\maketitle
\begin{abstract}
This paper separately evaluates the effects of inclination and asymmetry of solar coronal loops on the resonant absorption of kink magnetohydrodynamic (MHD) oscillations. We modelled a typical coronal loop by a straight and axisymmetric cylindrical magnetic flux tube filled with cold plasma. We solved the dispersion relation numerically for different values of the longitudinal mass density stratification. We show that, in inclined and asymmetric loops, the frequencies and their corresponding damping rates of the fundamental and first-overtone modes of kink oscillations are smaller in comparison with semi-circular uninclined loops with the same lengths. The results also indicate that, the period ratio $P_1/P_2$, increases with increasing the inclination of the loop, but it decreases less than $2\%$ while imposing the asymmetry to each loop side, up to $9.66\%$ of the loop length. The ratio of each mode frequency to  its corresponding damping rate remain unchanged approximately while the inclination or the asymmetry imposed. Hence, we conclude that these ratios are reliable for inferring the physical parameters of coronal loops and coronal medium, regardless of the loop shape or the state of its inclination. In addition, in contrast with the effect of asymmetry which is not significant on the period ratio $P_1/P_2$, when an observed oscillating loop has a smaller apex height, the state of its inclination is an important factor that should be considered, especially when the period ratio $P_1/P_2$, is taken into consideration for coronal seismology.
\end{abstract}


\section{Introduction}

In the recent years, the observations of the solar coronal structures with high resolution
instruments and high quality 3D techniques, make it possible to measure and infer important quantities of the coronal medium. Also modelling methods and magnetohydrodynamics (MHD) aided coronal seismology and granted valuable information in estimating directly-unobservable
physical properties of the solar corona. Since the first observation of transverse oscillations of coronal
loops by Aschwanden et al. (1999) and Nakariakov et al. (1999), a wealth of various investigations have
been made to explain and interpret these oscillatory behaviours and the mechanisms behind .

Verwichte et al. (2004) investigated transverse oscillations of nine loops in a coronal post flare arcade. For two of the loops, they observed two oscillation modes in each loop, simultaneously. The large period oscillation was approximately double the small period. They interpreted the long and short period oscillations as the fundamental mode and first overtone of kink oscillations, respectively. For the observed periodic phenomena, they reported the numbers $1.81 \pm 0.25$ and $1.64 \pm 0.23$ to be the ratio of the fundamental and the first-overtone kink mode periods $P_1/P_2$. Although those numbers reported by Verwichte et al. (2004), were later modified by Van Doorsselaere et al. (2007) to $1.82 \pm 0.08$ and $1.58 \pm 0.06$, respectively, however both sets of reported period ratios were obviously less than the canonical value $2$. The main reasons that have been proposed more recently to explain the departure of period ratios from their canonical values are longitudinal density stratification and magnetic field expansion (e.g. Andries et al. 2005; Karami \& Asvar 2007; Safari et al. 2006; Ruderman et al. 2016) or magnetic twist (see Erd\'{e}lyi \& Fedun 2006; Karami \& Bahari 2010; Karami \& Bahari 2012; Ebrahimi \& Karami 2016). Besides, in the presence of longitudinal density stratification some effects such as the deviation of loop shape from a semi-circular geometry can cause the oscillation of coronal loops to undergo some changes. For instance, Dymova \& Ruderman (2006b) considered a loop shape as an arc of a circle with radius $R$ in which, a parameter $\lambda$ determined how the center of the circle located with respect to the photosphere. They concluded that, the oscillation period ratio $P_1/P_2$, not only depends on the ratio of the loop height to the atmospheric scale height, but also on the parameter $\lambda$ considered for the loop. Therefore based on their results, the authors argued that, for a proper determination of scale height using observed value of $P_1/P_2$, it is important to know the shape of the loop.

Another important observational feature of the oscillations of solar coronal loops that has been reported by various studies (e.g. Nakariakov et al. 1999; Wang $\&$ Solanki 2004; Goddard et al. 2016), is their rapidly decaying behaviour. Up to now, several mechanisms such as damping due to viscosity and ohmic resistivity (e.g. Karami \& Barin 2009), wave leakage (e.g. De Pontieu et al. 2001), phase mixing (e.g. Heyvaerts \& Priest 1983; Ebrahimi et al. 2017) and resonant absorption (e.g. Andries et al. 2005; Safari et al. 2006; Pascoe et al. 2010) have been put forward to explain the damping of the oscillations. However among suggested mechanisms, resonant absorption are thought to be more likely at explaining the phenomena.

In an oscillating MHD structure such as a coronal loop, if the spatially varying Alfv\'{e}n frequency equals the global frequency somewhere inside the system, then resonant absorption will take place. Accordingly, the energy of global mode of the system transfers to the local Alfv\'{e}n waves in the vicinity of the resonance point which results to large gradients in the amplitudes of the perturbations in the resonance layer. In the presence of dissipation effects, this concentrated energy could be transformed to heat the plasma (e.g. Ebrahimi et al. 2020). Since Ionson (1978) who for the first time suggested resonant absorption as a plausible mechanism of heating for magnetic flux tubes in solar corona, various studies have been made on resonant absorption (e.g. Rae \& Roberts 1982; Davila 1987; Hollweg \& Yang 1988; Sakurai et al. 1991a,b; Steinolfson \& Davila 1993; Goossens et al. 1995; Ruderman \& Roberts 2002).
Van Doorsselaere et al. (2004a) studied resonant absorption in cylindrical magnetic flux tubes. They showed that, by increasing the width of the inhomogeneous layer of the tube, the results obtained from numerical calculations and those inferred from analytical approximation differ up to 25\%. Terradas et al. (2006a) solved a time- depended problem for an initial disturbance, and studied the induction of MHD kink wave in a coronal loop and the attenuation of the disturbance due to resonant absorption in thin and thick boundary layer regimes.

The effect of longitudinal mass density stratification on the resonant absorption of transverse oscillations of coronal loop, was first investigated by Andries et al. (2005). They showed that, in longitudinally stratified coronal loops, the frequency and the damping rate of fundamental kink wave increase as the stratification parameter increases. Also their results indicated that, the ratio of the damping rate to the frequency is independent of the stratification parameter. The results obtained by Andries et al. (2005) were later approved by Dymova \& Ruderman (2006a) even for arbitrary radial density profile in the inhomogeneous layer of the tube.
Karami et al. (2009) showed that in the presence of longitudinal density stratification the frequency ratio $\omega_2/\omega_1$, of resonantly damped oscillations is less than 2 for both kink and fluting modes that is confirmed by the observation for the kink waves but so far, the existence of fluting modes in the coronal flux tubes has not been reported. Their results indicated that, although both the frequencies and  their corresponding damping rates increase as the longitudinally stratification parameter increases; but the ratio of each frequency to its corresponding damping rate does not experience any change.

The deviation of the coronal loops shape from a customary semi-circular shape has been the topic of some studies in the recent decade. Morton $\&$ Erd\'{e}lyi (2009) showed how an elliptical shape of a coronal loop and its stage of emergence i.e., the position of its center with respect to photospheric surface can affect the kink oscillation period ratio $P_1/P_2$. Similarly Karami et al. (2013) studied the effect of an elliptic shape and the stage of emergence of longitudinally stratified coronal loops on the resonantly damped kink oscillations. They showed that both the degree of ellipticity and the stage of emergence of the loop alter the kink frequencies and damping rates of the tube as well as the ratio of frequencies of the fundamental mode and its first overtone. For the values of longitudinal mass density stratification parameter considered by Karami et al. (2013), their results for the period ratio $P_1/P_2$, were in agreement with the findings of Morton $\&$ Erd\'{e}lyi (2009).

The solar corona images taken by various methods reveal that the coronal loops are curved structures with their footpoints anchored in the photosphere. However, Van Doorsselaere et al. (2004b) and Terradas et al. (2006b) showed that, the effect of loop curvature is negligible on the loop oscillations. It implies that, one can consider any shape for a coronal loop by imposing the desired configuration to only the longitudinal mass density profile of the loop. One of the geometrical features of solar coronal loops that can play significant roles in hydrodynamic modeling and also interpreting coronal medium and loops seismology is the deviation of the loop plane from a plane normal to the photosphere. Aschwanden et al. (1999) analyzed three-dimensional structure of solar active region observed by SOHO and deduced some physical properties of 30 various coronal loops. They concluded that, the inclination angle of the loops covered a vast interval between $-56^{\circ}$ and $69^{\circ}$. Similarly, Reale et al. (2000) analyzed the data of a brightening coronal loop that had been observed with TRACE spacecraft. They concluded that, for a reasonable infer of loop geometry and stratification, an inclination angle of $\approx 60^{\circ}$ can be considered. Using the EUVI instrument of the two STEREO A and B spacecraft, Aschwanden et al. (2008) identified 30 coronal loop structures and computed their 3D coordinates. Their results showed that the inclination angle of the observed loops were in the range of $35.7^{\circ}$ to $72.8^{\circ}$. For a more recent study of analyzing coronal loop structures, see e.g. Aschwanden et al. (2011).

In an attempt to determine the true shape of coronal loops, Alissandrakis et al. (2008) reconstructed 3D shapes of two coronal loops observed in Ne VIII $770\AA$ and O V $630 \AA$ spectral lines by SOHO. They applied a method to calculate the flow velocity along coronal loops using the Doppler velocities and the true reconstructed shape of loops. The results of their investigation indicated that, in addition to large values of the inclination angles $55^{\circ}$ and $70^{\circ}$, the velocity profiles along the loops were asymmetric with respect to their apexes. Several observations in the past years have revealed some asymmetries in the spectral line of solar coronal loops. Doschek et al. (2007), Hara et al. (2008), De Pontieu \& McIntosh (2010) and Bryans et al. (2010) have made investigations in order to interpret the nature of and the mechanisms behind the asymmetric spectra of transition region or coronal medium structures. Orza \& Ballai (2013) attributed the asymmetrical nature of a coronal loop to its shape and studied MHD oscillations of asymmetric coronal loops in various degrees of asymmetricity. Their results indicated that, in the presence of longitudinal stratification of mass density, when the temperature is the same inside and outside the loop, the asymmetry can alter the kink wave period ratio $P_1/P_2$ by $5-8\%$.

 The present paper separately evaluates how the asymmetry of a coronal loop shape or the deviation of their planes from the vertical direction affect the resonant absorption of kink waves. In section 2 the governing equations and the models of the flux tube in two cases are introduced. In section 3 we derive the dispersion relation using the jump conditions. Section 4 is devoted to the numerical results. Finally, we conclude the paper in section 5.

\section{Equations of Motion and Modeling of the Flux Tube}\label{S2}
The linearized MHD equations for a cold plasma, neglecting the effect of gravity and in the absence of the background plasma flow are as follows
\begin{eqnarray} \frac{\partial\mathbf{v'}}{\partial
t}=\frac{1}{4\pi\rho}\{(\nabla\times\mathbf{B'})\times\mathbf{B}
+(\nabla\times\mathbf{B})\times\mathbf{B'}\}
+\frac{\eta}{\rho}\nabla^2\mathbf{v'},\label{mhd1}
\end{eqnarray}
\begin{equation}
\frac{\partial\mathbf{B'}}{\partial
t}=\nabla\times(\mathbf{v'}\times\mathbf{B})+
\frac{c^2}{4\pi\sigma}\nabla^2\mathbf{B'},\label{mhd2}
\end{equation}
with $\bf{v'}$ and $\bf{B'}$ being the Eulerian
perturbations of velocity and magnetic field, respectively. Also $\bf{B}$,
$\rho$, $\sigma$, $\eta$ and $c$ are the background
magnetic filed, the mass density, the electrical conductivity, the
viscosity and the speed of light, respectively. In Eqs. (\ref{mhd1}) and (\ref{mhd2}) the viscosity ($\eta$) and electric conductivity ($\sigma$) of the plasma are assumed to be constant everywhere.

As a typical coronal flux tube, we consider a straight cylinder with a circular cross section of radius $R$. We use the cylindrical coordinates ($r$, $\varphi$, $z$) in which the flux tube axis lies in the $z$ direction. The background magnetic field is constant throughout the medium and it is parallel to the tube axis, i.e., $\mathbf{B} = B_0\hat{z}$, where $B_0$ is a constant. We write the plasma density function of the medium as $\rho(r,z)=\rho_0(r)\rho(z)$ where the radial part is as follows
\begin{eqnarray}
\rho_0(r)&=&\left\{\begin{array}{ccc}
  \rho_{{\rm in}},&(r<R_1),&\\
  \Big [ \frac{\rho_{\rm in}-\rho_{\rm ex}}{l}\Big](R-r)+\rho_{\rm ex},&(R_1\leq r\leq R),&\\
   \rho_{{\rm ex}},&(r>R).&\\
   \end{array}\right.\label{rho0}
\end{eqnarray}
Here, $\rho_{\rm i}$ and $\rho_{\rm e}$ are the constant densities of the interior and exterior
regions of the flux tube, respectively and $l\equiv R-R_{\rm 1}$ indicates the thickness of the inhomogeneous layer.

The $t$- and $\varphi$- dependency of the perturbations are assumed to be $\exp\left[i(m\varphi -\tilde{\omega} t)\right]$ with $\tilde{\omega}=\omega-i\gamma$ being the complex frequency in which $\omega$ and $\gamma$ are the frequency and damping rate, respectively. Here, $m$ is the azimuthal mode number of the wave. Following Karami et al. (2009), we expand the perturbed quantities $\bf{v'}$ and $\bf{B'}$ in Eqs. (\ref{mhd1}) and (\ref{mhd2}) in terms of radial and longitudinal-dependent functions as
 \begin{eqnarray}\label{Brz}
\mathbf{B'}(r,z)=\sum_{k=1}^{\infty}\mathbf{B}^{(k)}(r)\psi^{(k)}(z),\nonumber\\
\mathbf{v'}(r,z)=\sum_{k=1}^{\infty}\mathbf{v}^{(k)}(r)\psi^{(k)}(z),
\end{eqnarray}
where the functions $\psi^{(k)}(z)$ form a complete set of orthonormal eigenfunctions of Alfv\'{e}n operator $L_{A}=\rho\left(\omega^2+v_A^2\frac{\partial^2}{\partial z^2}\right)$, and satisfy the  relation of eigenvalue problem $L_{A}\psi^{(k)}=\lambda_{k}\psi^{(k)}$ (see Andries et al. 2005). In the definition of Alfv\'{e}n operator $v_{A}=B_0/\sqrt{4\pi\rho}$ is the Alfv\'{e}n velocity.

In the absence of dissipation, solving Eqs. (\ref{mhd1}) and
(\ref{mhd2}) for body waves in the interior region $(r<R_1)$
leads to the solutions for the longitudinal component of $\bf{B'}$ and radial component of $\bf{v'}$ as follows (see Karami \& Asvar 2007):
\begin{eqnarray}
{B'}_z^{(\rm{in})}(r,z)=\sum_{{\rm k}=1}^{+\infty}A^{\rm{(in,k)}}J_{\rm
m}(|k_{\rm{in,k}}|r)\psi^{(\rm{in,k})}(z),\label{soli1}
\end{eqnarray}
\begin{eqnarray}
\begin{aligned}\label{soli}
{v'}_r^{(\rm{in})}(r,z)=-\frac{i\tilde{\omega}
B_0}{4\pi}\sum_{{\rm k}=1}^{+\infty}\Bigg[\frac{k_{\rm{in,k}}}{\lambda_{\rm{in,k}}}A^{(\rm{in,k})}\times J'_{\rm m}(|k_{\rm{in,k}}|r)\psi^{(\rm{in,k})}(z)\Bigg],\end{aligned}
\end{eqnarray}
in which $A^{(\rm in,k)}$ is a constant and $ \lambda_{\rm{ in,k}}$ is the eigenvalue of k'th mode of Alfv\'{e}n operator in the interior region and $k_{\rm{in,k}}^2=\frac{\lambda_{\rm{ in,k} }}{B^2/4\pi}$. Also, the prime on $J_m$ represents a derivative with respect to its argument. Substituting the modified Bessel function $K_m$ for the Bessel function of the first kind $J_m$, the index “ex" for the index “in", and $k_{\rm{ex,k}}^2=-\lambda_{\rm{ex,k} }/\left(B^2/4\pi\right)$ for $k_{\rm{in,k}}^2$  in Eqs. (\ref{soli1}) and (\ref{soli}) we obtain the expressions for the perturbations of the magnetic field and velocity in the external region.

\subsection{Inclined Loop}

We model an inclined coronal loop by a semi-circular magnetic flux tube for which the loop footpoints anchored in the solar photosphere. However we assume that, the loop plane is not in general perpendicular to the photospheric surface. Hence, let $\theta$ be the angle of deviation of loop's plane from a vertical plane normal to the photosphere. One can write the longitudinal mass density profile as (see e.g. Verth et al. 2008)
\begin{eqnarray} \rho(z)&=&\exp\Big[-\mu
\cos(\theta)\sin\big(\frac{\pi z}{L}\big)\Big ],\label{roz2}
\end{eqnarray}
 where the stratification parameter $\mu$ is defined in terms of the loop length $L$, and the density scale height $H$, as
 \begin{equation}\label{mu}
   \mu:= \frac{L}{\pi H}.
\end{equation}
 It is worth noting that, in this model, when the loop becomes inclined, its length remains constant. Therefore the loop apex height decreases with increasing the inclination angle. In order to the loop apex height remains constant maintaining its semi-circular shape the loop length must increase by increasing its inclination. Under that condition, the loop length in Eq. (\ref{mu}) must be replaced with $\frac{L}{\cos(\theta)}$. Substituting the relation into Eq. (\ref{roz2}), we see that the effects of inclination will be removed. In that case, only the effects of the loop length remain. In other words, the change in the frequency and damping rate is due to the change of the loop length not the inclination angle.

\subsection{Asymmetric Loop}
\begin{figure}
\centering
 \includegraphics[width=70mm]{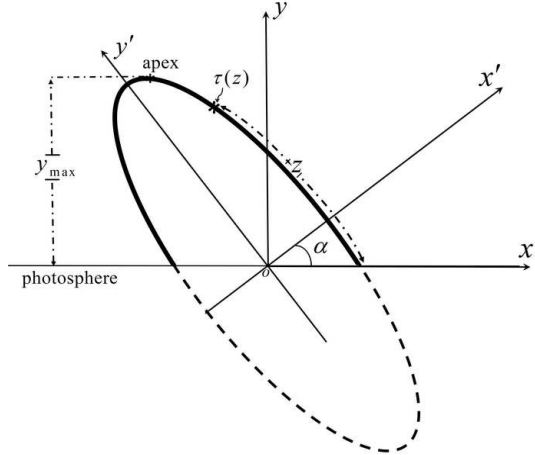}
\caption[] {A schematic representation of the geometry of an asymmetric coronal loop.}\label{eloop}
 \end{figure}

\begin{figure*}
\centering
\begin{tabular}{cc}
  \includegraphics[width=70mm]{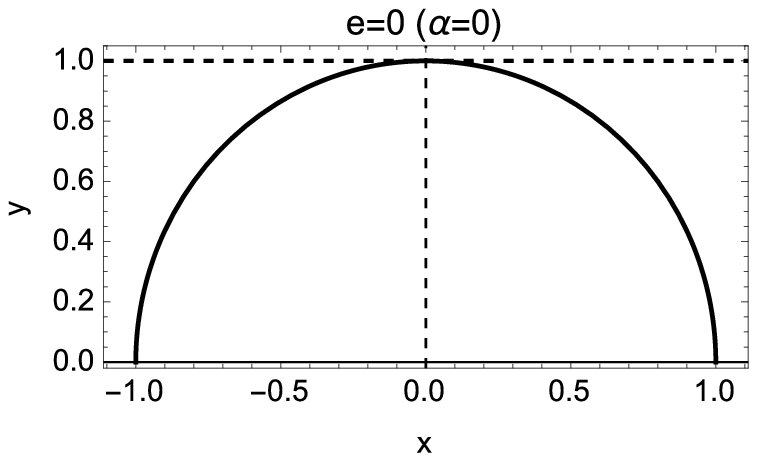}& \includegraphics[width=70mm]{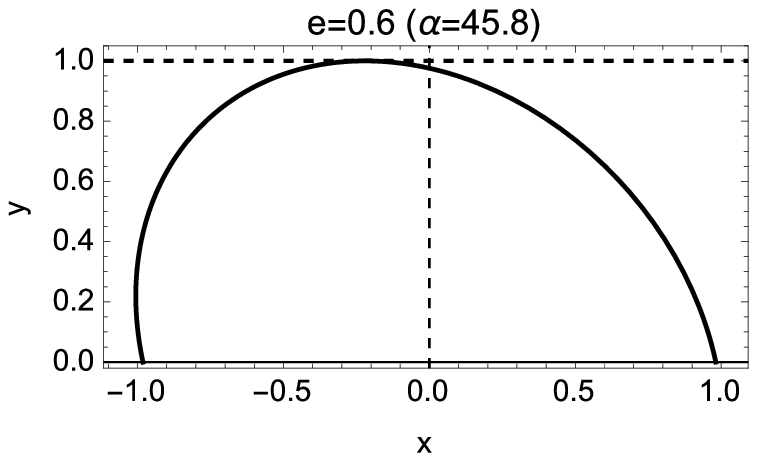} \\
  \includegraphics[width=70mm]{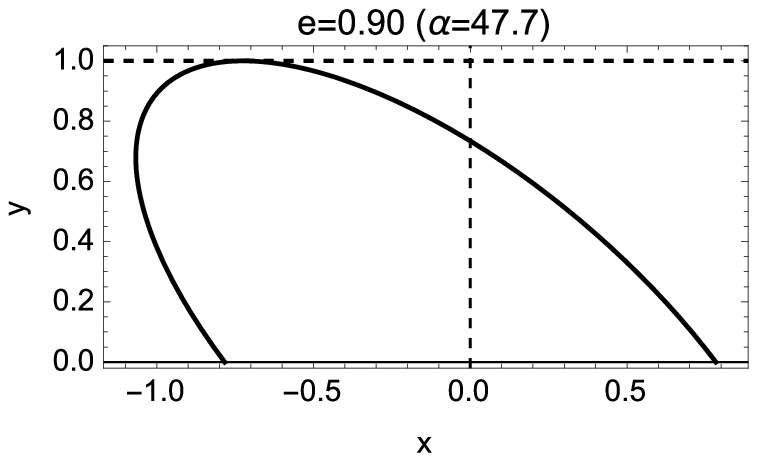}& \includegraphics[width=70mm]{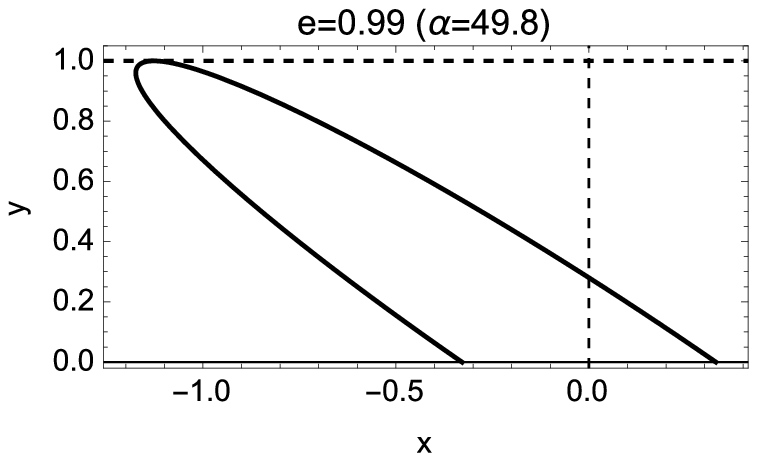} \\
\end{tabular}
\caption[] {Various stages of asymmetricity in the coronal loop. All lengths are normalized to the radius of curvature of the semi-circular loop.}\label{eloop1}
\end{figure*}

   In order to impose asymmetry to a modelled coronal loop, we must firstly deviate it from a common half-circle shape. Also we want the apex of the loop not to be necessarily the middle point of the loop length. Moreover the line joining the apex to the midpoint of loop footpoints distance does not lie in the vertical direction perpendicular to the photospheric surface (see Fig. \ref{eloop}). To this aim, we model the asymmetric loop as a part of an elliptical curve in which the asymmetricity arises firstly in the difference between the length of each side of the loop with respect to the apex point. Then as can be seen in Figs. \ref{eloop} and \ref{eloop1}, each side of the loop lies differently in the solar atmosphere, i.e., the left side which is situated in a lower case, rises with a larger slope from left footpoint to the apex, while the right side of the loop ascends with a smoother slope. Accordingly, in an asymmetric coronal loop modelled here, the details of longitudinal part of mass density profile will be different in the two sides of the loop. It is worth mentioning that, since we hope to merely show the effects of asymmetry in shape of coronal loops on their kink oscillation, the loop is modelled in such a way that two other geometrical parameters, the loop length and the apex height (see Fig. \ref{eloop1}), that may play roles in its oscillations characteristics, remain constant while applying the asymmetry. It should be noted that another motivation for the choice of such a model for an asymmetric loop is that the type of the asymmetry of the coronal loop we chose, includes a type of inclination of the loop itself. In the previous section we discussed a type of inclination of the loop in which a line joining the loop apex to the midpoint of the loop footpoints becomes inclined with the inclination of the loop plane. However, here in this model, the line is inclined, while the loop plane is still normal to the solar photosphere.

  As illustrated in Fig. \ref{eloop}, the major and minor half-axes of the modelled loop are coincide with the axes of the primed coordinate system $x'-y'$. The letter $\alpha$ that we use to denote the asymmetry parameter, is an angle through which the primed coordinate system is rotated with respect to the unprimed coordinate system $x-y$. The relation between the components of the two coordinate systems is given by the matrix relation
  \begin{equation}
  \left[ {\begin{array}{ccccccccccccccc}
 x \\
 y
\end{array}} \right] = \left[ {\begin{array}{ccccccccccccccc}
 {\cos \alpha }&{-\sin \alpha } \\
 {\sin \alpha }&{\cos \alpha }
\end{array}} \right]\left[ {\begin{array}{ccccccccccccccc}
 {x'} \\
 {y'}
\end{array}} \right].\label{rotateloop}
\end{equation}

   For a symmetric coronal loop the primed and unprimed coordinate system coincide $(\alpha=0)$, hence, in this case the loop is semi-circle. For any given $\alpha>0$, the loop shape is elliptic and somehow asymmetric. Also the loop ellipticity ($e$) increases with increasing the asymmetry parameter (see Fig. \ref{eloop1}). The mutual dependency between $e$ and $\alpha$ is determined through the constancy of loop apex height. The corresponding relation will be derived below.

   From Eq. (\ref{rotateloop}) we have
   \begin{equation}\label{y}
    y = x'\sin \alpha + y'\cos \alpha.
   \end{equation}
   Also one can write the components of the primed coordinate system in terms of a parametric angle along the loop as
   \begin{equation}\label{xyt}
  \begin{array}{l}
x' = a\cos\tau,\\
y' = b\sin\tau,
\end{array}
   \end{equation}
      with $a$ and $b$ being the minor and major half-axes of the elliptic loop. Now substituting Eq. (\ref{xyt}) in Eq. (\ref{y}) and then taking the first derivative with respect to $\tau$ yields
    \begin{equation}\label{dydt}
     \frac{{dy}}{{d\tau}} = b\cos \tau\cos \alpha - a\sin \tau\sin \alpha.
    \end{equation}
       To obtain the loop apex height with respect to the photosphere ($y_{\rm{max}}$), $dy/dr$ must be equal to zero. Hence, the value of parameter $\tau$ corresponding to the apex point is given by
    \begin{equation}\label{tapex}
     \tau_{\rm a} = \operatorname{Arctan} (\frac{b}{a}\cot \alpha ).
    \end{equation}
    Using Eq. (\ref{xyt}) and (\ref{tapex}) we get the height of the loop apex from Eq. (\ref{y}) as follow
    \begin{equation}\label{ymax}
     {y_{\max }} = a\sin \alpha \sqrt {1 + \frac{{{{\cot }^{2}}\alpha }}{{1 - {e^{2}}}}}~,
    \end{equation}
    where
    \begin{equation}\label{e}
     {e^2} = 1 - {\left( {\frac{a}{b}} \right)^2}.
    \end{equation}
         Let us recall that we use $r$ to indicate the radial coordinate in the loop cross-section, hence, to avoid ambiguity between it and radial coordinate of the polar coordinate system, we denote the latter by $r'$. Therefore the primed coordinates in terms of the polar coordinates $(r',\phi)$, are
    \begin{equation}\label{xypolar}
     \begin{array}{l}
     x' = r'\cos \phi, \\
     y' = r'\sin \phi,
\end{array}
    \end{equation}
    where the polar angle $\phi$ is measured with respect to the $x'$ axis.

    The equation of the ellipse in the primed coordinate system is
    \begin{equation}\label{ellipse}
     \frac{{{{x'}^2}}}{{{a^2}}} + \frac{{{{y'}^2}}}{{{b^2}}} = 1.
    \end{equation}
    Then substituting Eqs. (\ref{e}) and (\ref{xypolar}) in Eq.(\ref{ellipse}) leads to the relation for the radial coordinate
    \begin{equation}\label{r'}
    r' = \frac{a}{{\sqrt {1 - {e^2}{{\sin }^2}\phi } }}~.
    \end{equation}
     Now equating the right hand sides of Eqs. (\ref{xyt}) and (\ref{xypolar}) for $y'$, then using Eq. (\ref{r'}) yields
     \begin{equation}\label{tauphi}
      \sin \tau = \frac{{\sqrt {1 - {e^2}} \sin \phi }}{{\sqrt {1 - {e^2}{{\sin }^2}\phi } }}~,
     \end{equation}
     which is the relation between the parametric and polar angles.

\begin{figure}
\centering
 \includegraphics[width=70mm]{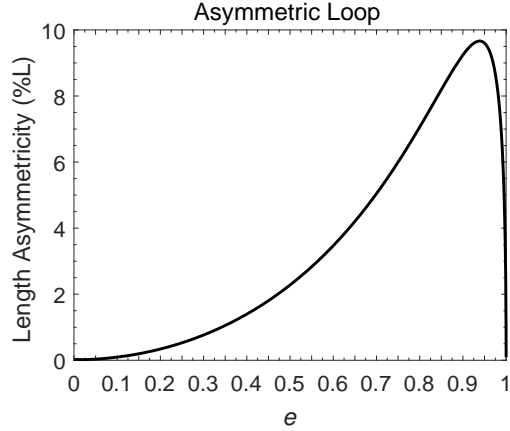}
\caption[] {Length difference between the two sides of the apex of an asymmetric loop versus the loop ellipticity. \label{lod}}
 \end{figure}

     According to Fig. \ref{eloop}, one can get the loop length in the polar coordinate system as
     \begin{equation}\label{L}
      L = \int_{ - \alpha }^{\pi - \alpha } {r'\sqrt{1+\frac{1}{{r'}^2}\left(\frac{dr'}{d\phi}\right)^2}d\phi }.
     \end{equation}
     Also using Eqs. (\ref{r'}) and (\ref{tauphi}) it is straightforward to obtain $L$ by integrating with respect to the parametric angle:
     \begin{equation}\label{L}
      L = \frac{a}{{\sqrt {1 - {e^2}} }}\int_{ - \operatorname{Arcsin} u}^{\pi - \operatorname{Arcsin} u} {\sqrt {1 - {e^2}{{\sin }^2}\tau } } d\tau~,
     \end{equation}
     where
     \begin{equation}\label{L1}
     u = \frac{{\sqrt {1 - {e^2}} \sin \alpha }}{{\sqrt {1 - {e^2}{{\sin }^2}\alpha } }}~.
     \end{equation}
     As shown in Fig. (\ref{eloop1}) for all values of the loop ellipticity, the loop apex height equals the radius of curvature of the semi-circle loop, i.e.,
     \begin{equation}\label{ymaxlp}
     y_{\rm{max}} = \frac{L}{\pi}.
     \end{equation}
     Now substituting Eq. (\ref{L}) in (\ref{ymaxlp}) and using it in Eq. (\ref{ymax}), gives
     \begin{equation}\label{ealpha}
        \int_{ - \operatorname{Arcsin} u}^{\pi - \operatorname{Arcsin} u} {\sqrt {1 - {e^2}{{\sin }^2}\tau } } d\tau-\pi \sin \alpha \sqrt {\rm{cosec}^2\alpha-\textit{e}^2} = 0.
     \end{equation}
        The expression is a constraint relation between $e$ and $\alpha$, i.e., once we assign a value to the loop ellipticity $e$, the value of asymmetry parameter is determined. According to the relation, taking the ellipticity of the loop in the interval $[0, 0.999]$ gives $\alpha \in[0, 50.3]$.
The constancy of loop length yields another constraint relation
\begin{equation}\label{a}
a = \frac{L }{\pi \sin\alpha} \sqrt {\frac{{1 - {e^{2}}}}{\rm{cosec}^2\alpha-\textit{e}^2}}~.
\end{equation}
This relation gives the minor half-axis length of the loop, for various values of $e$ and $\alpha$. Similar to Eq. (\ref{L}) one can obtain an arbitrary length along the loop axis as follow
\begin{equation}\label{z}
  z = \frac{a}{{\sqrt {1 - {e^2}} }}\int_{ - \operatorname{Arcsin} u}^{\tau(z)} {\sqrt {1 - {e^2}{{\sin }^2}\tau } } d\tau~,
\end{equation}
where $\tau(z)$ is the parametric angle corresponding to a distant $z$ from the right footpoint (Fig. \ref{eloop}). According to Eqs. (\ref{y}) and (\ref{xyt}) the height of a point on the loop from the photosphere is
\begin{equation}\label{y(z)}
  y(z) = \frac{a}{{\sqrt {1 - {e^2}} }}\left( {\sin \tau (z)\cos \alpha + \sqrt {1 - {e^2}} \cos \tau (z)\sin \alpha } \right).
\end{equation}\\
Now for the modelled asymmetric coronal loop, substituting Eq. (\ref{a}) in Eq. (\ref{y(z)}) and using the definition in Eq. (\ref{mu}), for the longitudinal part of plasma density
\begin{equation}\label{ro(z)2}
 \rho (z) = \exp \left( { - \frac{{y(z)}}{H}} \right),
\end{equation}
we have
\begin{equation}\label{ro(z)1}
  \begin{aligned}
    \rho (z) = \exp \left[-\frac{\mu }{\sqrt {\rm{cosec}^2\alpha -\textit{e}^2}}\left (\sin \tau (z)\cot \alpha  +\sqrt {1 - {e^2}} \cos \tau (z) \right) \right].
  \end{aligned}
 \end{equation}
Fig. \ref{lod} represents the asymmetry between the lengths of the two sides of the loop in percentage of the loop length $L$, versus the ellipticity of the loop $e$. According to the figure, when the ellipticity of the loop increases from 0 to $\sim0.94$, the length asymmetry also increases up to 9.66\% of $L$, i.e., 19.32\% of each side length in the half-circle loop. Increasing the ellipticity from $0.94$ to 0.99 leads to the decrease of length asymmetry with a steeper slope. Note that, in presence of asymmetry, different sides of the loop have different slopes in rising from the footpoint to the loop apex (see Fig. \ref{eloop1}). Hence, we expect the longitudinal density profile to has different slope in each side. For a more clear explanation let denote the right and left sides of the loop by ``ri'' and ``le'' respectively. It is easy to find out that
 \begin{equation}\label{lefright}
 \frac{d\rho(z)}{dy(z)}|_{\rm{ri}}=\frac{d\rho(z)}{dy(z)}|_{\rm{le}}.
 \end{equation}
 Using the chain rule yields
 \begin{equation}\label{lefright}
 \frac{d\rho(z)}{dz}\frac{dz}{dy(z)}|_{\rm{ri}}= \frac{d\rho(z)}{dz}\frac{dz}{dy(z)}|_{\rm{le}}.
 \end{equation}
 The above expression means that, in an equal height from the photosphere, the side with a greater raising slope (see Fig. \ref{eloop1}), also has a greater longitudinal density slope along the loop.

\section{Connection Formulae and Dispersion Relation}\label{sec3}
According to Eq. (\ref{rho0}), in the transitional layer $(R_1<r<R)$, the plasma density decreases
from $\rho_{\rm{in}}$ to $\rho_{\rm{ex}}$, continuously. The radial inhomogeneity of the plasma density results to a continuum of Alfv\'{e}n speeds in the radial direction. The necessary condition for the occurrence of resonant absorption mechanism is that at a particular radial distance $r_A$, the frequency of the global kink mode equals the local Alfv\'{e}n frequency (Ionson 1978). Consequently, the amplitudes of oscillatory perturbations diverge in a thin layer around the resonance point and the energy of the global kink wave transfers to the transitional layer of the flux tube. Accordingly, since the dissipative effects are no longer negligible in the resonance layer, the equations of motion (\ref{mhd1}) and (\ref{mhd2}) must be fully solved. Up to now, it seems to be a very arduous task to find any analytical solution for the complete form of these equations in the resonance layer. However, Sakurai et al. (1991a) (see also Goossens et al. 1995) suggested relations by which one can overcome the difficulty and pass the resonance layer needless to know the solutions in the layer. In fact, these relations provides the jumps of the analytical solutions across the resonance layer. The jump conditions are valid when the thickness of the dissipative layer is much smaller than the thickness of the transitional layer. Hence, to obtain the expression for the jumps of any quantity through the transitional layer we need to solve the ideal MHD equations in the transitional layer at the two sides of the dissipative layer and then use the connection formulae to connect the two solutions. However, the damping rate due to resonant absorption is defined by the jump of the imaginary part of the radial velocity. Jump of this quantity across the transitional layer is equal to its jump across the dissipative layer. The jumps of the total pressure perturbation and the real part of the radial plasma displacement only provide a small correction to the oscillation frequency. If we disregard this correction, then there is no need to solve the ideal MHD equations in the transitional layer far from the resonant surface. Hence, taking the jumps across the transitional layer equal to the jumps across the dissipative layer, we still obtain the correct expression for the decrement. Similar to Karami et al. (2009, 2013) it follows from Thompson \& Wright (1993) that the jump relations across the inhomogeneous layer read
 \begin{eqnarray}
\left[B'_z\right]= 0,\label{jumps}
\end{eqnarray}
\begin{eqnarray}
 \left[v'_r
 \right]=-\sum^{+\infty}_{k=1}\frac{B_0\tilde{\omega} m^2\left\langle {{\psi_0^{\rm_{(in,k)}}}}
 \mathrel{\left| {\vphantom {{\psi_0^{\rm_{(in,k)}}} {{B'}_z^{\rm_{(in,k)}}}}}
 \right. \kern-\nulldelimiterspace}
 {{{B'}_z^{\rm{(in,k)}}}} \right\rangle}{4r_A^2\left\langle {{\psi_0^{\rm{(in,k)}}}}
 \mathrel{\left | {\vphantom {{\psi_0^k} {\L_{A1}\psi_0^{\rm{(in,k)}}}}}
 \right. \kern-\nulldelimiterspace}
 {{L_{A1}\big |\psi_0^{\rm{(in,k)}}}} \right\rangle}\psi_0^{\rm{(in,k)}},\label{jumps1}
 \end{eqnarray}
where $L_{A1}$ denotes the derivative of $L_A$ at Alfv\'{e}n
radius
\begin{eqnarray}
 L_{A1}=\tilde{\omega}^2[1+\sum_{n=1}^{+\infty}C_{n}S_{n,k,k}]\frac{\partial\rho_0(r)}{\partial
r}\Big|_{r=r_A},\label{La1}
\end{eqnarray}
with
\begin{equation}\label{cn}
 {C_n} = \int_0^L {\Big( \rho (z) - 1 \Big)} \sin ({\textstyle{{n\pi } \over L}}z)dz~~~~~~~~~~~~n\geq 1
\end{equation}
and
\begin{eqnarray}
S_{n,k,j}=\frac{2}{L}\int_0^L\sin\left(\frac{n\pi}{L}
z\right)\sin\left(\frac{k\pi}{L}
z\right)\sin\left(\frac{j\pi}{L} z\right){\rm
d}z,
 \end{eqnarray}
also $\psi_0^{\rm_{(in,k)}}$s being the eigenfunctions
of $L_A$ with vanishing eigenvalues ($\lambda_{k}=0$). The jump of each quantity in Eqs. (\ref{jumps}) and (\ref{jumps1}) is equal to its relevant contribution in the exterior region minus that in the interior region. If one substitute the fields of Eqs. (\ref{soli1}) and (\ref{soli}) in jump relations, it yields to a matrix equation. It is straightforward to show that the matrix equation have non-trivial solution if the infinite matrix in the equation has vanishing determinant, i.e.

\begin{eqnarray}
 \left| {\begin{array}{*{20}c}
  {\Pi_1^{(\rm ex,1)} } & {-\Pi_1^{(\rm in,1)}} & {\Pi_1^{(\rm ex,2)} } & { - \Pi_1^{(\rm in,2)} } & \ldots  \\
  {\Xi_1^{(\rm ex,1)} } & {\Lambda_1^{(\rm in,1)}} & {\Xi_1^{(\rm ex,2)} } & {\Lambda_1^{(\rm in,2)} } & \ldots  \\
  {\Pi_2^{(\rm ex,1)} } & {-\Pi_2^{(\rm in,1)}} & {\Pi_2^{(\rm ex,2)} } & { - \Pi_2^{(\rm in,2)} } & \ldots  \\
  {\Xi_2^{(\rm ex,1)} } & {\Lambda_2^{(\rm in,1)}} & {\Xi_2^{(\rm ex,2)} } & {\Lambda_2^{(\rm in,2)} } & \ldots  \\
  \vdots & \vdots & \vdots & \vdots & \ddots  \\
\end{array}}\right|
= 0,\label{disperssion}
\end{eqnarray}
where
\begin{align}\label{dispe}
  &\Lambda_j^{(\rm in,k)}=\Xi_j^{(\rm in,k)} + {\mathcal{D}}_j^{(\rm in,k)},\nonumber\\
  &\Pi_j^{({\rm{in}},k)}={J_m}({x_k})\psi _j^{({\rm{in}},k)}, \nonumber \\
  &\Pi_j^{({\rm{ex}},k)}={K_m}({y_k})\psi _j^{({\rm{ex}},k)}, \nonumber\\
  &\Xi_j^{({\rm{in}},k)}=\frac{{{{J'}_m}({x_k})}}{{{x_k}}}\psi _j^{({\rm{in}},k)}, \nonumber\\
  &\Xi_j^{({\rm{ex}},k)}=\frac{{{K_m}({y_k})}}{{{y_k}}}\psi _j^{({\rm{ex}},k)},
\end{align}
\begin{equation}\label{dispe1}
 \mathcal{D}_j^{({\rm{in}},k)} = - i\frac{{{B_0^2}{m^2}l\sum\limits_{n = 1}^{ + \infty } {{\psi _0}_n^{({\rm{in}},k)}\Pi _n^{({\rm{in}},k)}} }}{{4{R^3}{{\tilde \omega }^2}({\rho _{\rm{ex}}} - {\rho _{\rm{in}}})(1 + \sum_{n=1}^{+\infty}C_{n}S_{n,k,k})}}{\psi _0}_j^{({\rm{in}},k)},
\end{equation}
with
\begin{eqnarray}
 x_k &=& \left| {{k_{{\rm{in}},k}}R} \right|,\nonumber \\
 {y_k} &=& \left| {{k_{{\rm{in}},k}}R} \right|,\nonumber \\
 \psi _j^{({\rm{in}},k)} &=& \left\{
  \begin{array}{l}
  (\frac{L}{\pi R})^2\frac{\tilde\omega^2\sum_{n=1}^{+\infty}C_{n}S_{n,k,j}}{j^2-k^2},{\rm{~~~~~~~~~~~~}}k \ne j,\\
  1,{~~~~~~~~~~~~~~~~~~~~~~~~~~~~~~~~~~~~~~~~~~~~~~~~~~}k = j,
  \end{array} \right.\nonumber \\
 \psi _j^{({\rm{ex}},k)} &=& \left\{ \begin{array}{l}
(\frac{\rho_{\rm{ex}}}{\rho_{\rm{in}}})^2(\frac{L}{\pi R})^2\frac{\tilde\omega^2\sum_{n=1}^{+\infty}C_{n}S_{n,k,j}}{j^2-k^2},{\rm{~}}k \ne j,\\
1,{~~~~~~~~~~~~~~~~~~~~~~~~~~~~~~~~~~~~~~~~~~~~~~~~~~}k = j,
\end{array} \right.\nonumber \\
 \psi_{0j}^{({\rm{in}},k)} &=& \left\{ \begin{array}{l}
{\textstyle{{{{k}^2}{\sum_{n=1}^{+\infty}C_{n}S_{n,k,j}}} \over {\rho_{\rm{in}}(1+\sum_{n=1}^{+\infty}C_{n}S_{n,k,k})({j^2} - {k^2})}}},{\rm{~~~~}}k \ne j,\\
1,{~~~~~~~~~~~~~~~~~~~~~~~~~~~~~~~~~~~~~~~~~~~~~~~~~~}k = j,
\end{array} \right.
\end{eqnarray}
The infinite system in Eq. (\ref{disperssion}) is the dispersion relation which solving it, gives the frequencies $\omega$,
and the damping rates $\gamma$, of kink ($m=1$) oscillations of the loop. It can be shown that, if one reduces the expression to determinants with finite orders, and then solves each finite dispersion relation numerically, there exists a value, say $\rm{n}_{\rm{min}}$, for the order of the finite determinant, such that, taking determinants with orders larger than $\rm{n}_{\rm{min}}$, does not change the results, significantly (see Andries et al. 2005). The reason is that the entries $S_{k,j} = S_{n,k,j}$ are dominant only along $\pm n$th off diagonal, where $n$ is the larger index in dominant $C_n$s in Eq. (\ref{cn}). We see that, taking a determinant of order 10, is accurate enough to give a good approximation of the frequencies and damping rates.

\section{Numerical Results}
To obtain the frequencies and damping rates of kink oscillations, we solve the dispersion relation (\ref{disperssion}) numerically. The parameters of the modelled coronal loop are $R/L=0.01$ , $l/R=0.02$, $\rho_{\rm{ex}}/\rho_{\rm{in}}=0.1$. We normalized all obtained frequencies and damping rates with respect to
$\omega_{A_{\rm{in}}}:= \frac{v_{A_{\rm{in}}}}{L}$. We note that, since the transitional layer is thin, we take $r_A\approx R$ in the dispersion relation wherever it is needed.

It is worth noting that although we have assumed a thin flux tube ($R/L=0.01$) in our model but the method of Andries et al. (2005) that we have used in this paper is still valid when the tube is not thin. The only assumption in this method is thin boundary layer ($l/R\ll 1$) approximation. Dymova \& Ruderman (2005, 2006a) developed an alternative simpler method that is only valid in the thin tube approximation.

In what follows we present the numerical results for two separate cases of inclined and asymmetric loop.

\subsection{Inclined Loop}
\begin{figure}
\centering
 \includegraphics[width=70mm]{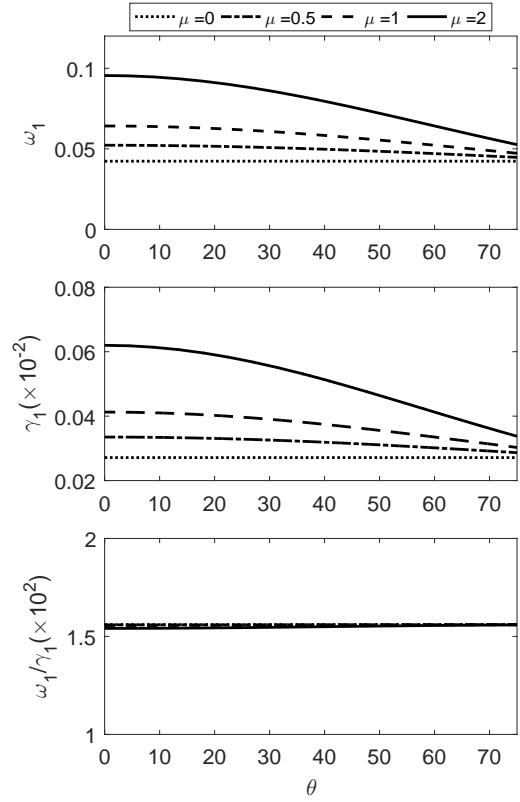}

\caption[] {The oscillation frequency (upper panel),  the
damping rate (middle panel)
 and the ratio of the frequency and  the damping rate (bottom panel) versus the loop plane inclination angle $\theta$, for the fundamental kink modes.
 The loop parameters are $R/L=0.01$, $l/R=0.02$,
$\rho_{\rm{ex}}/\rho_{\rm{in}}=0.1$ and $e=0$.
\label{inc1}}
 \end{figure}

\begin{figure}
\centering
 \includegraphics[width=70mm]{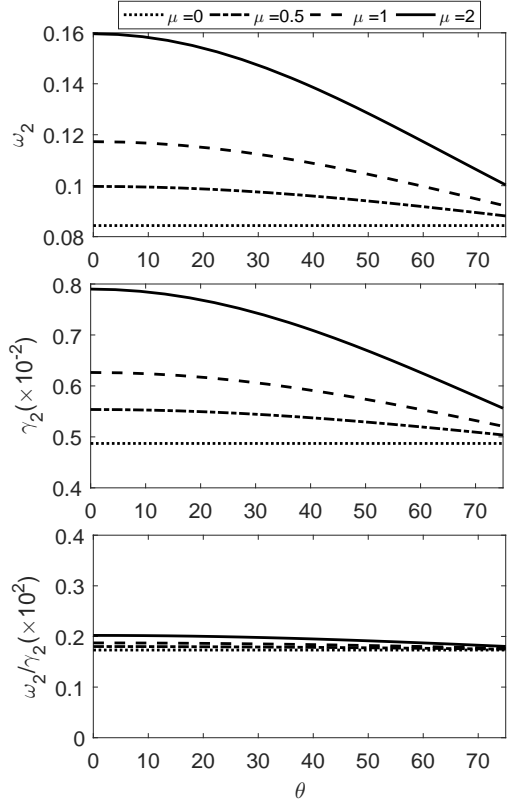}

\caption[] { Same as Fig. \ref{inc1} but for the first-overtone of kink oscillation.\label{inc2}}
 \end{figure}

\begin{figure}
\centering
 \includegraphics[width=70mm]{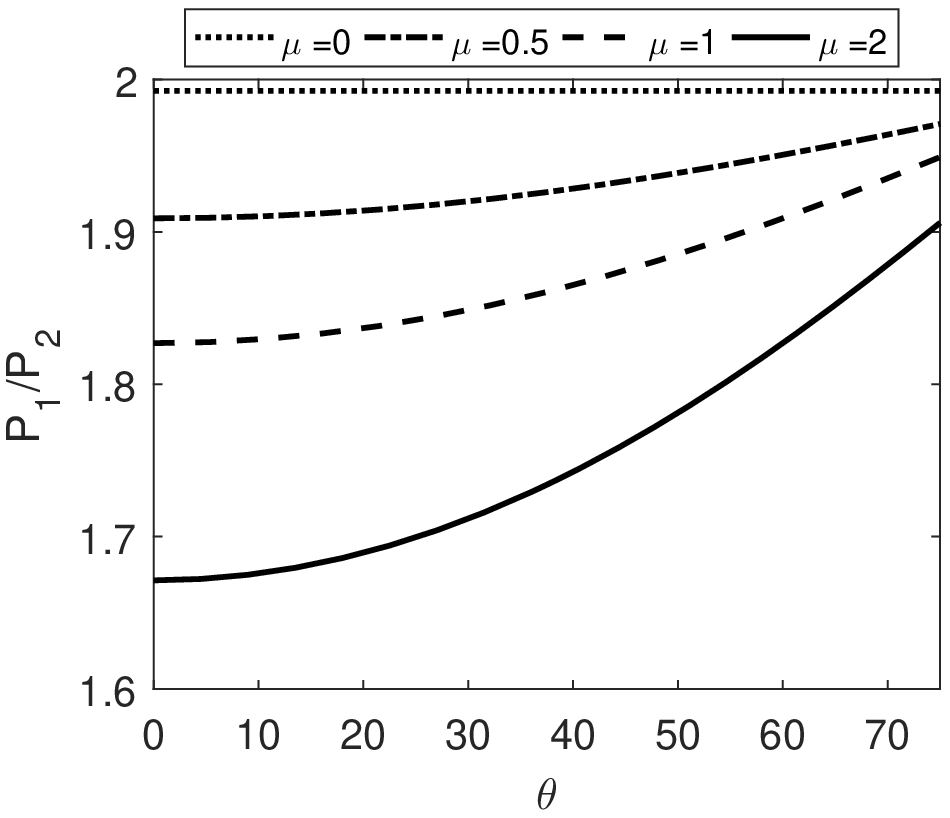}
\caption[]{Period ratio of the fundamental mode to the first-overtone of kink oscillation versus the loop plane inclination angle
$\theta$. The parameters are the same as in Fig. \ref{inc1}.
\label{inc3}}
 \end{figure}

Figures \ref{inc1}, \ref{inc2} and \ref{inc3} represent the results of numerical computations for both fundamental mode and first overtone of kink oscillations in longitudinally stratified inclined loops. In Figs. \ref{inc1} and \ref{inc2} we have plotted the frequency $\omega$, the damping rate $\gamma$, and the ratio $\omega/\gamma$ , versus the loop plane inclination angle $\theta$, for four values of longitudinal stratification parameter $\mu=(0, 0.5, 1, 2)$. In the upper and middle panels of the figures we see that, in longitudinally stratified loops, as the inclination angle of the loop plane increases from $0$ to $75^{o}$ , the frequencies ($\omega_1$, $\omega_2$) and the damping rates ($\gamma_1$, $\gamma_2$) decrease. Also the figures reveal that, the rate of the variations of the frequencies and the damping rates depend on the longitudinal density stratification parameter. In other words, the higher the value of $\mu$, the greater the rate and the amount of the variation are. For $\mu=2$, as the inclination angle increases from  $0$ to $75^{o}$, $\omega_1$, $\gamma_1$, $\omega_2$ and $\gamma_2$ decrease by $44.9\%$, $45.5\%$, $37.1\%$ and $36.2\%$, respectively. For uninclined loop, the frequencies and the damping rates increase with increasing the longitudinal density stratification parameter (Andries et al. 2005; Karami et al. 2009, 2013). According to  Figs. \ref{inc1} and \ref{inc2} this pattern is also valid for inclined loops, but by increasing the inclination angle, the differences between the frequencies and the damping rates for different values of $\mu$ decrease and are going to be zero in a very rare or unrealistic inclination angle $\theta=90^{o}$.

The bottom panels of Figs. \ref{inc1} and \ref{inc2} represent the ratios $\omega_1 /\gamma_1$ and $\omega_2 /\gamma_2$ versus the inclination angle for a symmetric loop, i.e. $e=0$, respectively. The ratio of each mode frequency to its corresponding damping rate, indicates the number of oscillations before complete damping. It is clear from the figures that, for all values of longitudinal density stratification parameter $\mu$, the ratios are independent of the inclination angle and approximately the same for all values of $\mu$ as has been shown previously for uninclined loops (see Andries et al. 2005; Dymova \& Ruderman 2006a; Karami et al. 2009). Therefore, the ratio is a reliable quantity to estimate directly unobservable quantities such as $l/R$ and density contrast $\rho_{\rm{ex}}/\rho_{\rm{in}}$ (see Dymova \& Ruderman 2006a), regardless of the degree of longitudinal density stratification and the state of inclination.

In Fig. \ref{inc3} the ratio of periods of the fundamental mode and the first-overtone $P_1/P_2$ of a symmetric loop, i.e. e=0, is plotted versus the inclination
angle for four values of longitudinal density stratification parameter. Figure shows that, for longitudinally stratified loops ($\mu>0$), when the loop plane inclination angle increases, the period ratio increases as well. Also the figure shows that the rate and amount of the period ratio variation, depends on the value of $\mu$, i.e., the higher the degree of stratification, the greater the rate and the value of period ratio variation are. For $\mu=2$ as the inclination angle $\theta$ increases from $0$ to $75^{o}$ the period ratio increases by 14\%. In addition, Fig. \ref{inc3} reveals that although the period ratio decreases by increasing $\mu$, but the difference between the values of period ratio for different values of longitudinal stratification parameter, decrease when the inclination angle increases. However, this result is expectable, since as the loop become more inclined, its height reduces and therefore the effect of longitudinal density stratification decreases. These results indicates that, for a comprehensive interpreting of coronal seismology using period ratio, the inclination of the loop plane is one of the important factors that should be took into account during loop oscillation observations, especially when stronger longitudinal density stratifications are considered.

 \subsection{Asymmetric Loop}
\begin{figure}
\centering
\includegraphics[width=70mm]{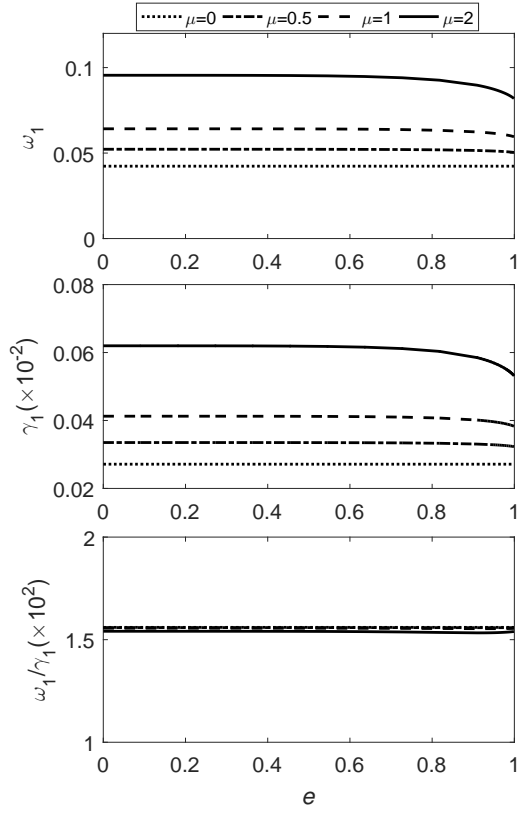}
\caption[] {The oscillation frequency (upper panel),  the damping rate (middle panel) and the ratio of the frequency and  the damping rate (bottom panel) versus the asymmetric loop ellipticity $e$, for the fundamental kink mode. The loop plane inclination angle is zero, i.e. $\theta=0$. Other parameters are the same as in Fig. \ref{inc1}.
\label{asym1}}
 \end{figure}

\begin{figure}
\centering
 \includegraphics[width=70mm]{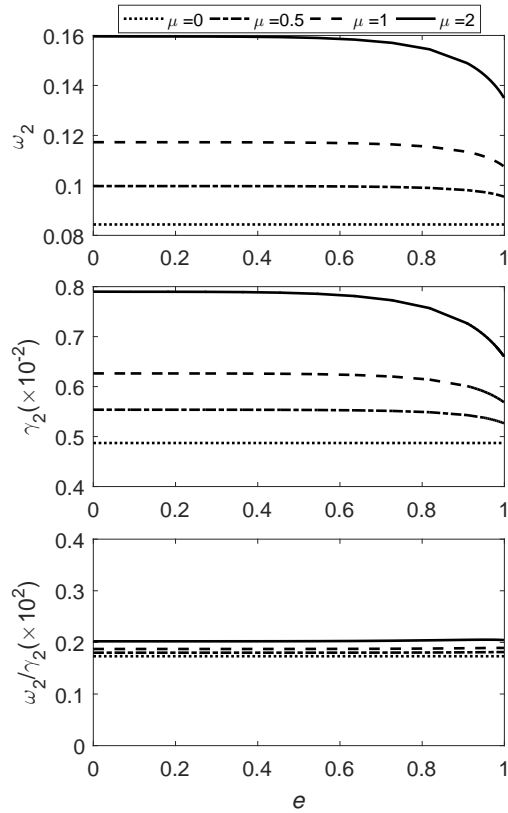}
\caption[] {Same as Fig. \ref{asym1} but for the first-overtone.\label{asym2}}
 \end{figure}

\begin{figure}
\centering
 \includegraphics[width=70mm]{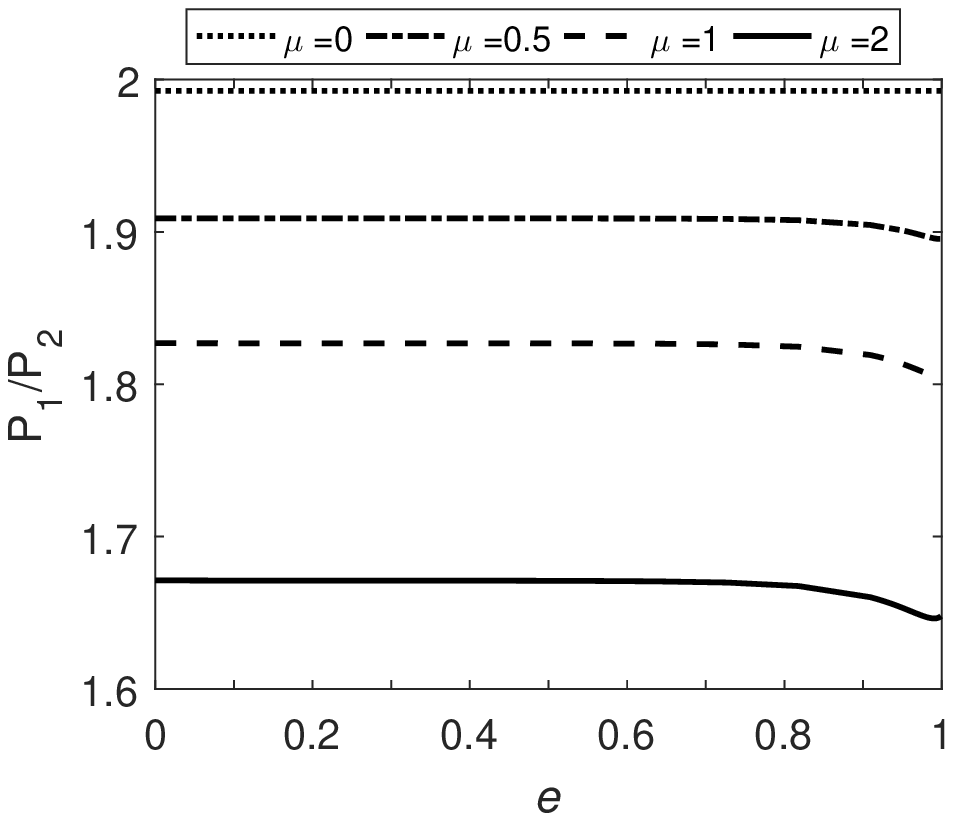}
\caption[]{Period ratio of the fundamental mode to the first-overtone versus the asymmetric loop ellipticity $e$. The parameters are the same as in Fig. \ref{inc1}.
\label{asym3}}
 \end{figure}

  Figures \ref{asym1}, \ref{asym2} and \ref{asym3} illustrate the effects of asymmetry of coronal loops on the resonant absorption of fundamental mode and first overtone of kink oscillations for four values of longitudinal density stratification parameter $\mu=(0, 0.5, 1, 2)$ with $\theta=0$. In the upper and middle panels of Figs. \ref{asym1} and  \ref{asym2} the frequencies ($\omega_1, \omega_2$) and the damping rates ($\gamma_1, \gamma_2$) are plotted with respect to the loop ellipticity $e$. The figures reveal that, for all values of $\mu$, as the ellipticity increases from 0 up $\sim 0.55$, the frequencies and damping rates remain approximately unchanged. But for longitudinally stratified cases $(\mu>0)$, increasing the ellipticity from 0.55 to 0.999 leads to the decrease of the frequencies and the damping rates. Also it is clear from the figures that the amount and rate of the decrease of the frequencies and the damping rates are directly proportional to the degree of longitudinal stratification. For $\mu=2$, increasing the ellipticity of the loop up to $0.999$ leads to decrease of $\omega_1$, $\gamma_1$, $\omega_2$ and $\gamma_2$ by 14.2\%, 14.1\%, 15.4\% and 16.4\%, respectively. However it seems that the asymmetry between the length of the two sides of the loop substantially affects the frequencies and damping rates if it leads to the difference between the longitudinal density slopes ($d\rho(z)/dz)$) in the two sides of the loop. In fact, a justification for the constancy of the frequencies and damping rates on the interval $e\in[0,0.55]$ is that the length difference between the two sides (up to $\sim 0.03 L$) is not large enough to make a significant difference between the longitudinal density slopes in the two sides of the loop at equal heights from the photosphere. On the other hand, for the values of the loop ellipticity larger than $\sim 0.94$ despite the decrease of the length difference between the two sides, the decreasing pattern of the frequencies and the damping rates continues. It likely occurs due to the increase of difference between longitudinal density slopes in the two sides of the loop (see Fig. \ref{eloop1}).

 The bottom panels in Figs. \ref{asym1} and \ref{asym2} represent the ratios $\omega_1/\gamma_1$ and $\omega_2/\gamma_2$ against the loop ellipticity $e$, respectively. It is clear from the figures that changing the ellipticity on the interval $[0, 0.999]$ does not change the ratios. Therefore, imposing an asymmetry to the loop up to 9.66\% $L$, thereupon imposing asymmetry to the longitudinal density slopes of the two sides of the loop, has no significant effect on the number of the oscillations before complete damping. Hence if the longitudinal density stratification is uniform and the same inside and outside the loop, then the ratios $\omega_1/\gamma_1$ and $\omega_2/\gamma_2$ are equal for symmetric and asymmetric loops with the same apex height. It implies that this ratios could be used for deducing coronal loops quantities using coronal seismology regardless of the shape and the degree of asymmetry of the loop .

 Fig. \ref{asym3} represents the ratio of periods of fundamental mode and first-overtone of kink oscillation $P_1/P_2$ versus the ellipticity of the asymmetric loop. The figure reveals that for all values of $e\in [0, 0.999]$, by increasing the longitudinal density stratification parameter $\mu$, the period ratio decreases. According to the figure, the period ratio is approximately constant for the values of the ellipticity smaller than $\sim 0.55$. However, its variation for $e>0.55$ is not significant, even for the loops with larger values of $\mu$. For $\mu=2$, as $e$ increases from 0 to 0.999, the period ratio decreases by $1.4\%$. Hence for the longitudinal density stratification parameter value up to 2, the asymmetry of the loop shape up to $9.66\%$ of its length has no significant effect on the period ratio. It is worth noting that, although the variation of the period ratio is small, but the graphs for $\mu>0$ in Fig. \ref{asym3} are strictly decreasing even when the length difference between the two sides decreases(see Fig.\ref{lod}). It means that the variation of the period ratio is more due to the difference between $d\rho(z)/dz$ in the two sides rather than the length difference itself. Orza \& Ballai (2013) obtained 5\% for the change of kink wave period ratio $P_1/P_2$, for the longitudinal stratification parameter $L/\pi H =2$, and a parameter denoting the loop apex displacement $\alpha=0.6$ . However, since the loop length varies in their model (see page 3 in Orza \& Ballai 2013), the difference between the lengths and also the longitudinal density slopes of the two sides of the loop can reach the values much greater than those of our model. For instance, in the model of Orza \& Ballai (2013) for $\alpha=0.6$, the loops length changes by 21\% in comparison with a semi-circular loop, and the length difference between the two sided is 46.7\% of the loop length. Accordingly it shows that a large degree of asymmetry is needed in the coronal loops to affect the period ratio by a few percents.

\section{Conclusions}\label{sec5}
This paper aimed to illustrate that how the inclination of a coronal loop plane and the asymmetry in its shape can affect the resonant absorption of kink oscillations. In both cases we modelled a coronal loop with an axisymmetric flux tube embedded in a uniform background magnetic filed. In the radial direction the mass density linearly decreases across an inhomogeneous layer from the constant value inside the loop to the constant value outside. In the longitudinal direction along the loop axis, the density is stratified. Based on the method introduced by Andries et al. (2005), we solved the equations of motion in the absence of dissipative effects. Then using connection formulae, we derived the relevant dispersion relation and solved it numerically for the fundamental mode and the first-overtone.

We dealt with  the inclination of a semi-circle coronal loop in two cases. In the first case, as the loop becomes inclined, the loop apex height remains constant. Therefore, in this case, the loop length must increase while imposing the inclination to the loop. It can be shown that, in this case, the effects of inclination is removed and, as a result, the changes in the frequencies and damping rates are not due to the inclination of the loop, but they occur due to the change of the loop length. In contrast to the first case, in the second case the loop length remains constant while imposing the inclination to the loop. Therefore, in the second case, the apex height of the inclined loop must be less than that of an uninclined loop. Our results revealed that, in this case, depending on the degree of longitudinal mass density stratification, and the degree of inclination of the loop plane, the frequencies and damping rates as well as the period ratio $P_1/P_2$ can change significantly. For the longitudinal mass density stratification parameter $\mu=2$, as the inclination angle of loop plane increases from 0 up to 75$^{o}$:\\
(i) The frequencies $\omega_1$ and $\omega_2$ decrease by 44.9\% and 37.2\% respectively.\\
(ii)  the damping rates $\gamma_1$ and $\gamma_2$, decrease by 45.5\% and 36.2\% respectively.\\
(iii) The ratio $\omega/\gamma$ is approximately unchanged for both modes.\\
(iv) The period ratio $P_1/P_2$ increases by 14\%.\\
Hence, the results indicate that the inclination of the plane of an observed coronal loop is an important factor that must be took into account, especially while inferring physical condition of the coronal medium using coronal loops oscillations.

We modelled an asymmetric coronal loop by an elliptic arc in which the footpoints are anchored in the photosphere. By increasing the ellipticity of the loop the asymmetry between the two sides of the loop can reach 9.66\% of the loop length. The asymmetry is imposed to the loop in such a way that its length as well as its apex height remain constant. It enabled us to evaluate the effects of the asymmetry more clearly. For $\mu=2$, as the ellipticity of the loop increases form 0 up to 0.999:\\
(i) the frequencies $\omega_1$ and $\omega_2$ decrease by 14.2\% and 15.4\%, respectively.\\
(ii) absolute values of the damping rates $\gamma_1$ and $\gamma_2$, decrease by 14.1\% and 16.4\%, respectively.\\
(iii) the ratio of the frequency to its corresponding damping rate $\omega/\gamma$ is approximately unchanged for both modes.\\
(iv) the period ratio $P_1/P_2$ decrease only by 1.4\%.\\
Therefore, these results reveal that, although the frequencies and damping rates can change under the effect of asymmetry in highly longitudinally stratified loops, but the period ratio $P_1/P_2$ does not change considerably.
Also it is worthwhile to stress that the asymmetry between the length of the two sides of the loop only affects the loop oscillations if it causes to the difference between the longitudinal mass density slopes $d\rho/dz$ of the two sides. One reason for this claim is that, even when the length difference between the two sides decreases, the frequencies, damping rates as well as the period ratio continue to change in the same manner as before. A reliable explanation for this behaviour is the increase of the difference between the longitudinal mass density slope of the two sides.

In this study, we considered two cases of inclined and asymmetric loop. In the first case the loop is still semi-circular and the line joining the loop apex to the midpoint of the loop footpoints is inclined with the inclination of the loop plane. Whereas in the second case where the loop is no longer semi-circular, the line is inclined while the loop plane is still normal to the photosphere. As a common result for the two cases studied, we conclude that any type of the inclination of the coronal loop and asymmetry between the two sides that preserve the loop apex height has no significant effect on the kink oscillation period ratio $P_1/P_2$, whether the loop length remain constant or not.

According to the results, the ratio $\omega/\gamma$ for the fundamental mode and first-overtone of kink oscillations, is not affected by either the inclination or the asymmetric shape of the loop. Dymova \& Ruderman (2006a) showed analytically that, in semi-circular uninclined loops, this ratio depends only on the density contrast $\rho_{\rm{ex}}(z)/\rho_{\rm{in}}(z)$ and the ratio of the inhomogeneous layer thickness to the loop cross-section radius, $l/R$. Our results revealed that, when the stratification inside and outside the loop is still the same, the ratio $\omega/\gamma$ is not affected by the configuration of longitudinal density in inclined or asymmetric loops. It implies that, the ratio is a reliable quantity to use, for inferring physical parameters of coronal medium and coronal loops, regardless of the shape of the observed loop or the degree of its inclination.

\section*{Acknowledgements}
The authors thank the anonymous referee for very valuable comments.



\end{document}